\documentclass[pre,superscriptaddress,amsmath,amssymb,nofootinbib,floatfix]{revtex4-1}
\usepackage{amsfonts, amsmath, amssymb, amsthm, array, eucal, mathtools, xfrac}
\usepackage{booktabs, array, color}
\usepackage{natbib}
\newlength\fwidth
\newlength\fheight
\usepackage{graphicx}
\usepackage{siunitx}

\begin{document}

\title{Flow analysis of the low-Reynolds number swimmer \textit{C. elegans}}
\author{Thomas D. Montenegro-Johnson}
\thanks{These authors contributed equally to this work.}
\affiliation{Department of Applied Mathematics and Theoretical Physics, Centre
for Mathematical Sciences, University of Cambridge, Wilberforce Road, Cambridge
CB3 0WA, UK}
\author{David A. Gagnon}
\thanks{These authors contributed equally to this work.}
\affiliation{Department of Mechanical Engineering and Applied Mechanics,
University of Pennsylvania, Philadelphia, PA 19104}
\author{Paulo E. Arratia}
\email{parratia@seas.upenn.edu}
\affiliation{Department of Mechanical Engineering and Applied Mechanics,
University of Pennsylvania, Philadelphia, PA 19104}
\author{Eric Lauga}
\email{e.lauga@damtp.cam.ac.uk}
\affiliation{Department of Applied Mathematics and Theoretical Physics, Centre
for Mathematical Sciences, University of Cambridge, Wilberforce Road, Cambridge
CB3 0WA, UK}
\date{\today}
\begin{abstract}
Swimming cells and microorganisms are a critical component of many biological
processes. In order to better interpret experimental studies of low Reynolds
number swimming, we combine experimental and numerical methods to perform an
analysis of the flow-field around the swimming nematode \textit{Caenorhabditis
elegans}. We first use image processing and particle tracking velocimetry to
extract the body shape, kinematics, and flow-fields around the nematode. We then
construct a three-dimensional model using the experimental swimming kinematics
and employ a boundary element method to simulate flow-fields, obtaining very good
quantitative agreement with experiment. {We use this numerical model
to show that calculation of flow shear rates using purely planar data results in
significant underestimates of the true three-dimensional value. Applying symmetry arguments,
validated against numerics, we demonstrate that the out-of-plane contribution can
be accounted for via incompressibility and therefore simply calculated from particle tracking velocimetry.}
Our results show how fundamental fluid mechanics considerations may be used to
improve the accuracy of measurements in biofluiddynamics.
\end{abstract}

\maketitle

\section{Introduction}

The study of microorganism and cell swimming has numerous applications in both
industry and medicine, for instance in the context of mammalian
reproduction~\citep{johnson1999beltsville, Fauci2006}, biofuel
production~\citep{bees2014mathematics}, and the design of artificial biomedical
systems~\citep{qiu2015magnetic}. Many of these swimmers, such as mammalian
spermatozoa, self-propel by generating traveling undulations along their body
\cite{Lauga2009,gaffney11}. One such undulatory swimmer is the biological model
organism \textit{Caenorhabditis elegans}, a multi-cellular, free-living slender
nematode worm found in soil environments. 

The genetics and physiology of this nematode are well-studied. Its genome has been
completely sequenced~\cite{Brenner1974} and a complete cell lineage has been
established~\cite{byerly1976}.  The  neuromuscular system of \textit{C.~elegans}
controls its body undulations which allow it to swim, dig, and crawl through
diverse environments. The wealth of available biological knowledge thus makes
\textit{C.~elegans} an ideal candidate for investigations that combine aspects
of biology, biomechanics, and the fluid mechanics of propulsion. 

Recently, \textit{C.~elegans} has been used extensively as a model system for
experimental studies of propulsion, particularly at low Reynolds number, due to
its simple planar swimming gait and
size~\cite{gray1964,korta2007,ryu2014,shen2011undulatory,bilbao2013,
gagnon2014undulatory,park2016}. The nematode generates planar bending waves
through contractions of its ventral and dorsal muscles, producing a
quasi-two-dimensional (2D) traveling sinewave along its
body~\cite{korta2007,sznitman2010,thomases2015mechanisms}. At around
$1\,\mathrm{mm}$ in length and 75~\si{\micro}m in diameter, \textit{C.~elegans}
is significantly larger than the majority of low-$\mathrm{Re}$ undulatory
swimmers, enabling high-resolution reconstruction and analysis of planar flow
fields from particle tracking data. The resulting flow-fields can be used to
probe properties of both the swimmer and fluid, providing new insights into the
physics of undulatory propulsion~\citep{shen2011undulatory,
gagnon2013undulatory,ryu2014,shelley2012,bilbao2013,
gagnon2014undulatory,bau2015,park2016}.

\begin{figure}[b!]
    \centerline{\includegraphics{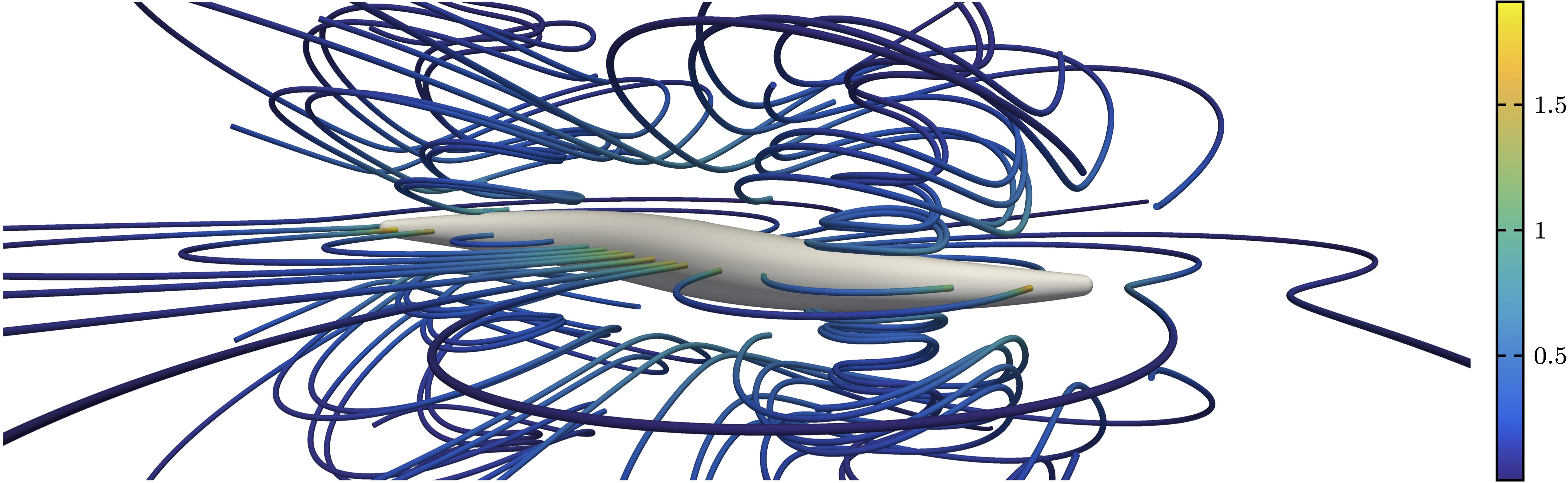}}
    \caption{Instantaneous flow streamlines genetrated by \textit{C.~elegans},
	computed numerically from experimental waveform data. The streamlines
    show a complex, three-dimensional flow-field. Color bar: speed $\mathrm{mm}/\mathrm{s}$.}
    \label{fig:c_elegans_3d}
\end{figure}

However, despite exhibiting a planar swimming stroke, the flow around
\textit{C.~elegans} has a complex three-dimensional structure
(Fig.~\ref{fig:c_elegans_3d}) which is difficult to capture experimentally. Many
useful flow properties, such as the shear rate, are dependent upon out-of-plane
flow contributions not measured by planar velocimetry data; in essence, one
measures a 2D slice of a three-dimensional (3D) field. This
limits, for example, our ability to accurately determine drag and propulsion
forces experienced and exerted by the swimming nematode. In this work, we show
that numerical and theoretical fluid mechanics techniques can be used to improve
the processing and analysis of these experimental flow-fields. 

We first obtain detailed (experimental) imaging data on the shape and kinematics
of a swimming \textit{C.~elegans}, together with particle-tracking data for
flow-field reconstruction. We then develop a numerical boundary element model of
the nematode, with geometry and boundary conditions specified directly from the
experiments, and use this model to improve the processing of particle tracking
velocimetry results. Using simulated flow-fields, we examine the spatial
distribution of flow shear rate, an important quantity related to power
dissipation and relevant for studies on locomotion in complex, shear-dependent
fluids. We show that using purely planar data significantly underestimates the
true value of the shear rate {throughout the field. We then use
symmetry arguments validated against numerics to show that the 2D measurements
can be corrected for out-of-plane effects by applying the incompressibility
constraint.} Our work shows that fundamental fluid mechanics tools can be used
alongside experimental measurements to improve our understanding of the
biomechanics of locomotion. 

\section{Methods}

We first describe the technical improvements carried out in this
work. In order to compare the results of the numerical model directly with our
experimental data, we use the same swimming movie to construct experimental
velocity fields and to obtain the motion of the nematode, which provides the
nematode's geometry and kinematics for the numerical simulations. 

Particle tracking procedures around swimmers entail a delicate balance between
maximizing the quantity of statistics from which to construct a velocity field,
and the precision of those measurements. For \textit{C.~elegans}, one needs
enough statistics to construct a smooth, differentiable velocity field, yet the
most precise particle tracking data is located strictly at the nematode
$z$-mid-plane (direction across the depth of field). Additionally, due to the
fact that our microscope objectives have a depth of field of approximately
20~\si{\micro}m, our experiments represent a \textit{depth-averaged} 2D slice of
a 3D flow-field over this thickness. 

In this paper, simulation provides benchmark flow-fields, allowing greater
selectivity for our particle identification and tracking algorithms. This
selection ensures that we only choose particles close to the $z$-mid-plane,
shifting the balance from maximizing the number of statistics to maximizing the
precision of our measurements. In regions where we have limited data, we use
gentle maximum value and bilateral filtering algorithms to smooth our
experimental results so that they can be differentiated. 

\begin{figure}[tb]
    \centerline{\includegraphics{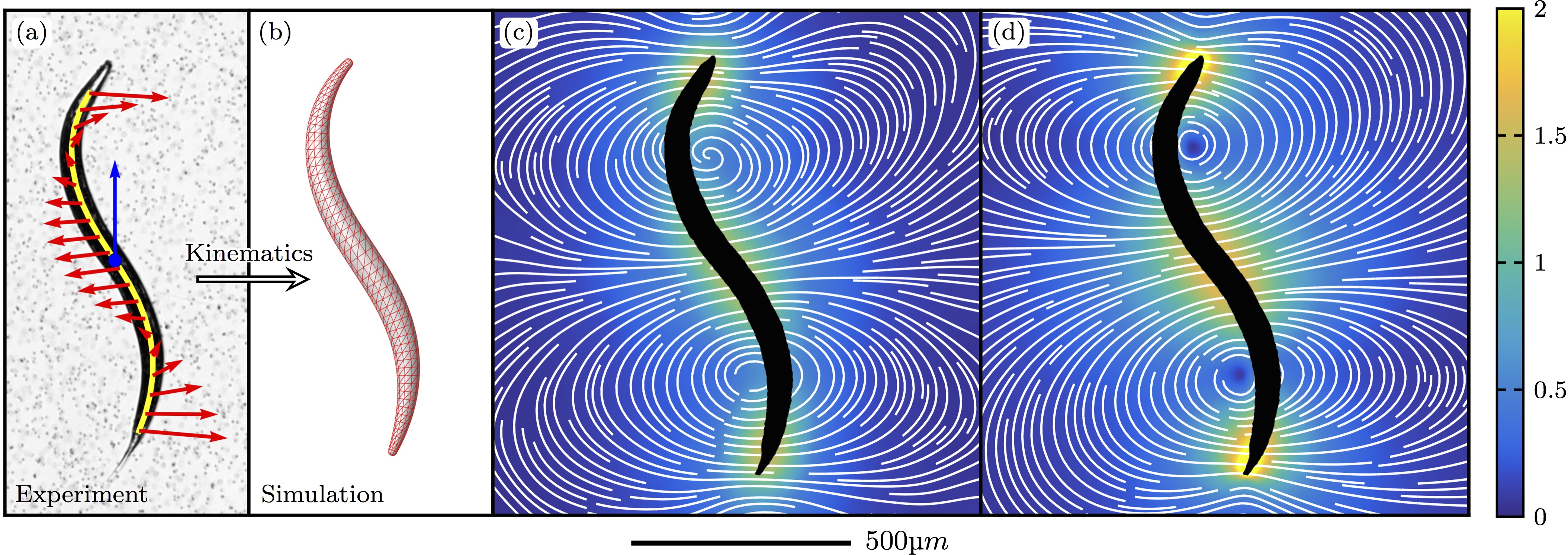}} 
    \caption{\textit{(a)} Image of the nematode \textit{C.~elegans}. The
        yellow line shows the ``skeletonized'' body, and the red arrows
        denote the velocities of each body-segment. The blue arrow
        indicates the swimming direction.
        \textit{(b)} Three-dimensional mesh constructed from the
        experimentally-obtained body-shape and kinematics of the nematode for
        the same snapshot. \textit{(c)} Experimentally-measured streamlines for
        one snapshot of swimming \textit{C.~elegans}, produced using particle
        tracking velocimetry techniques. \textit{(d)} Flow-field produced using
        the reconstructed 3D mesh and a numerical model for the same snapshot.
    Color bar: speed $\mathrm{mm}/\mathrm{s}$.} 
 \label{fig:method}
\end{figure}

\subsection{Data acquisition and processing}

Experiments with \textit{C.~elegans} are performed in Newtonian, water-like
($\mu \approx 1 \,\mathrm{mPa}\cdot·\mathrm{s}$ and $\rho \approx 10^3 \,
\mathrm{kg}/\mathrm{m}^3$) M9 buffer solutions~\citep{Brenner1974} in a sealed
fluid chamber 20~mm in diameter and 1~mm in depth. { Images are captured using standard bright-field microscopy (Infinity K2/SC microscope with a CF-4 objective, and an IO Industries Flare M180 camera at 150 frames per second). The depth of focus of the objective is approximately 20~\si{\micro}m, and we ensure that the focal plane is in the middle of the chamber in order to minimize wall effects.  The nematode beats primarily in the observation plane; the out-of-plane beating amplitude is less than 6\% of its in-plane motion, and therefore confinement effects are minimal~\citep{Sznitman2010PoF}. Given that the flow decay is expected to be exponential, we also anticipate negligible influence of the chamber boundaries on the flow field (see Fig.~\ref{fig:decay})~\citep{lighthill1976flagellar}.} All data presented here pertain to nematodes swimming at the centre of the fluidic chamber and out-of-plane recordings are discarded to avoid nematode-wall interactions and to minimize three-dimensional flow effects.

In-house software is used to track the swimming motion of \textit{C.~elegans}. The
position of the nematode's centroid is differentiated with respect to time to
obtain swimming speed ($U \approx 0.3$~mm/s), the head position is used to
compute an average amplitude ($A \approx 0.25$~mm), and a comparison of periodic
body shapes is used to estimate the nematode's beating frequency ($f \approx
2$~Hz). Beyond average kinematic properties, the nematode body contour is
automatically extracted for each image frame, skeletonized (yellow line in
Fig.~\ref{fig:method}\textit{a}), and divided into segments that are tracked to
obtain local body velocities. Sample velocity vectors of the body are shown in
Fig.~\ref{fig:method}\textit{a} in red. This body shape and associated
velocities are used to create an approximate 3D model of the nematode and its
surface velocities, as illustrated in Fig.~\ref{fig:method}\textit{b}. 

Particle tracking velocimetry is used to measure the velocity fields generated by swimming \textit{C.~elegans}. In short, we seed the
chamber with 3~\si{\micro}m polystyrene tracer particles
(Fig.~\ref{fig:method}\textit{a}) that are tracked using with in-house algorithms. We track the flow
for approximately 8 beat-cycles, and divide each cycle into approximately 50
phases. Because the nematode's swimming stroke is highly periodic ($\sim$ 2~Hz)~\citep{Sznitman2010PoF},
we can construct a ``master'' swimming cycle using a least-squares fit of the
nematode's body shapes. This phase-averaging technique considerably improves the
spatial resolution of the experimental velocity fields. In order to estimate the
flow-field around the nematode's body more accurately, we include velocities
within the contour of the the swimmer-fluid interface, assuming a no-slip
boundary condition. Finally, the data points for each phase, including tracking
plus boundary conditions, are averaged into gridded spaces of size
13.2~\si{\micro}m. The resulting experimental streamlines are illustrated in
Fig.~\ref{fig:method}\textit{c} for one particular phase. 

\subsection{Numerical model}
In the experimental setup, \textit{C.~elegans} moves through a water-like
Newtonian buffer solution \citep{Brenner1974}.  Thus, the characteristic
Reynolds number for the flow, $\mathrm{Re} = \rho U L/\mu$, is less than unity
($\mathrm{Re} \leq 0.3$). As such, the dynamics of the flow driven by the
swimming nematode is well-modeled by the Stokes flow equations 
\begin{equation} 
    \mu\nabla^2\mathbf{u} -\nabla p = 0,
    \quad \nabla\cdot\mathbf{u}=0, 
    \label{eq:stokes_flow}
\end{equation} 
where $\mathbf{u}$ is the fluid velocity and $p$ the dynamic pressure.

In order to solve Eqs.~\eqref{eq:stokes_flow} in the fluid surrounding the
nematode, we employ the regularized stokeslet boundary element
method~\citep{Cortez05,Smith09b}. The velocity throughout the domain is given by
integrals of stokeslets $\mathbf{S}$ and stresslets $\mathbf{T}$ over the
nematode's surface, $S$
\begin{align}
    \lambda u_j(\mathbf{x}_0) = \int_S
    &S_{ij}^\epsilon(\mathbf{x},\mathbf{x}_0)f_i(\mathbf{x})
    \nonumber \\
    &-u_i(\mathbf{x})T_{ijk}^\epsilon(\mathbf{x},\mathbf{x}_0)
    n_k(\mathbf{x})\,\mathrm{d}S_x,
    \label{eq:reg_bem}
\end{align}
for unknown surface tractions, $\mathbf{f}$, and surface velocity $\mathbf{u}$
specified from experimental data. 

{Note that in the majority of implementations of the method of
    regularised stokeslets, the stresslet ``double-layer''
    $u_i(\mathbf{x})T_{ijk}^\epsilon(\mathbf{x},\mathbf{x}_0)n_k(\mathbf{x})$
    term in equation~\eqref{eq:reg_bem} is eliminated, and the constant $\lambda
    = 1$~\citep{pozrikidis1992boundary}. However the surface tractions
    solved for in the simplified ``single layer'' formulation are a modified
    force density dependent upon a fictitious ``complementary'' flow inside the
    worm. In order to provide a general method that might in the future be used
    to examine the energetics of locomotion and force generation inside the
worm, the full formulation is used.}

In our case, the constant $\lambda$ is given at leading
order by $ \lambda\approx 1/2+ \kappa\epsilon/4$, where $\kappa$ is the mean
local curvature of the surface at $\mathbf{x}_0$~\citep{montenegro2015regularised}. 
We use the regularized form of the stokeslet $S_{ij}^\epsilon$ and stresslet
$T_{ijk}^\epsilon$~\cite{Cortez05},
\begin{subequations}
\begin{align}
    S_{ij}^\epsilon(\mathbf{x},\mathbf{x}_0) &= \frac{\delta_{ij}(r^2 +
    2\epsilon^2) + r_i r_j}{r_\epsilon^3}, \\
    T_{ijk}^\epsilon(\mathbf{x},\mathbf{x}_0) &= -\frac{6r_i r_j
    r_k}{r_\epsilon^5} \nonumber \\
    &\phantom{={ }}-\frac{3\epsilon^2\left(r_i\delta_{jk} + r_j\delta_{ik} + 
    r_k\delta_{ij}\right)}{r_\epsilon^5},
\end{align}
\end{subequations}
derived from a regularisation of the dirac delta function of the form 
\begin{equation}
    \phi_\epsilon(\mathbf{x} -
    \mathbf{x}_0) = \frac{15\epsilon^4}{8\pi r_\epsilon^7}, \quad r_\epsilon^2 =
    r^2 + \epsilon^2,
    \label{eq:blob_choice}
\end{equation}
where $r_i = (\mathbf{x} - \mathbf{x}_0)_i, r = |\mathbf{x} - \mathbf{x}_0|$ and
$\epsilon \ll 1$ (with $\epsilon=10^{-4}L$ in our simulations). The implementation uses routines adapted from BEMLIB~\citep{pozrikidis2002practical}
and the authors' boundary element library RegBEM Phoretic~\citep{monjohn}, which
employs a linear panel representation of the unknown surface tractions
$f_i(\mathbf{x})$ with adaptive Fekete quadrature for near-singular element
integrals~\citep{montenegro2015regularised}. 

\begin{figure}[tb]
    \centerline{\includegraphics{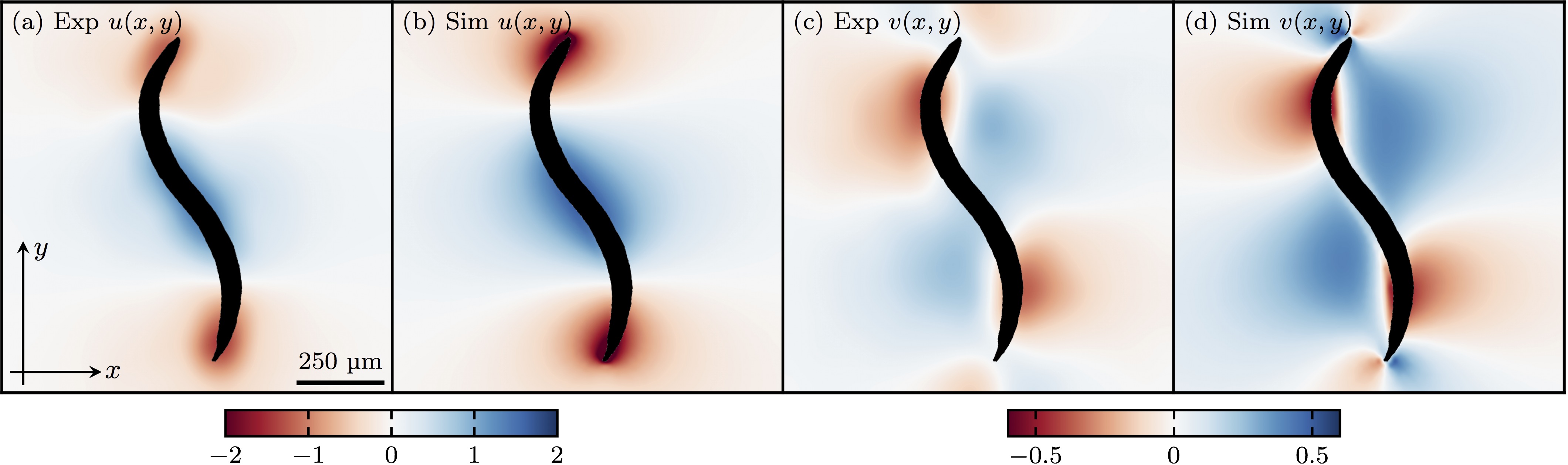}}
    \caption{Spatial distribution of the (a) experimentally-measured $x$-component $u(x,y)$, (b) simulated $x$-component, (c) 
experimentally-measured $y$-component $v(x,y)$, and (d) simulated $y$-component. Experimental measurements
demonstrating good agreement with simulation.
 Color bars: component velocity magnitude $\mathrm{mm}/\mathrm{s}$.}
    \label{fig:vel_comps}
\end{figure}

The nematode geometry is meshed with piecewise-quadratic triangular elements
(Fig.~\ref{fig:method}{\it b}) using a custom routine which extrudes a circle of
radius $a(s)$ along the experimentally-captured nematode centerline, and caps the
head and tail of the worm with a section of a sphere. The radius $a(s)$ is given
as a function of arc-length $s$ by fitting a quadratic through the nematode's radius
at the midpoint $a_1$ and the head $a_2$, so that for a nematode of length $L$ we
have
\begin{equation}
    a(s) = \frac{4(a_2 - a_1)}{L^2}\left(s - \frac{L}{2}\right)^2 + a_1.
\end{equation}
The velocity boundary condition (no-slip) is then imposed on the mesh surface
directly from experimental data via time-centered finite differences of the
nematode centerline. 

\section{Results and Discussion}
\subsection{Comparison of experiments and simulations}

\begin{figure}[tb]
    \centerline{\includegraphics[width=\textwidth]{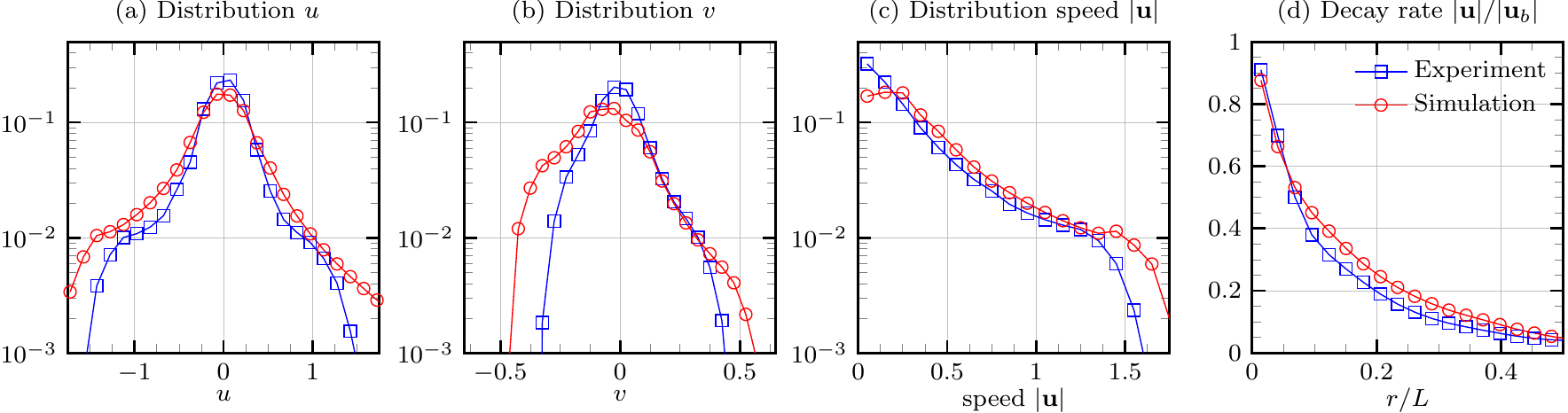}}
    \caption{{ Comparison of flow structure between experiment and
    simulation: (a) distribution of $u$, (b) distribution of $v$, (c)
distribution of speed $|\mathbf{u}|$, and (d) decay rate $|\mathbf{u}|/|\mathbf{u}_b|$.}}
    \label{fig:decay}
\end{figure}

Fig.~\ref{fig:method} shows the streamlines computed experimentally using
particle tracking velocimetry \textit{(c)} and numerically using simulated
flow-fields \textit{(d)}. These streamlines show strong agreement, with both
methods capturing head and tail vortices of similar shape and size. A visual
comparison of the $x$- and $y$-components of the velocity field, $u$ and $v$
respectively, again shows very good agreement between experiment and simulation
(Fig.~\ref{fig:vel_comps}). {In order to quantify this comparison,
Fig.~\ref{fig:decay}\textit{(a,b,c)} shows the distributions of velocity
components $u$ and $v$ and speed $|\mathbf{u}|$ for experimental and simulated fields.}
These plots show that the experiments capture the majority of velocities, with
the exception of the highest velocities corresponding to points closest to the
nematode, where it is difficult to obtain accurate particle tracks. Experiments
also predict a larger proportion of zero velocities, which is again associated
with the difficulty of extracting smooth, small, but non-zero, velocities from
noisy particle data.

Finally, we calculate the spatial decay of the flow speed away from
the body of the swimmer. For an undulatory swimmer, we expect to observe an
exponential flow decay~\citep{lighthill1976flagellar},
\begin{equation}
    \frac{\left|\mathbf{u} \right|}{ \left| \mathbf{u}_b \right|} = \exp\left(-{\frac{2 \pi r}{\alpha L}}\right),
\label{decay}
\end{equation}
where $\left| \mathbf{u} \right|$ is a velocity measurement, $\left| \mathbf{u}_{b} \right|$ is
the speed of the swimmer's body, assuming a no-slip boundary
condition, $r$ is the normal distance from each velocity measurement to the
swimmer's body, $L$ is a characteristic length-scale of the swimmer (we take
$L\approx 1$~mm), and $\alpha$ is an exponential parameter that specifies the
rate of decay. We find that the simulations and experiments are in good
agreement with $\alpha_{\mathrm{exp}} = 0.74$ and $\alpha_{\mathrm{sim}} =
0.85$, showing a difference of just 13\% in the exponent (Fig.~\ref{fig:decay}, last panel).

\subsection{Shear rate calculation and correction}

A particularly salient characteristic of the flow-field is the spatial
ditribution of shear rate, {$\dot{\gamma} =
\sqrt{\boldsymbol{\varepsilon}(\mathbf{u}):\boldsymbol{\varepsilon}(\mathbf{u})/2}$
with $\boldsymbol{\varepsilon}(\mathbf{u}) = \boldsymbol{\nabla}\mathbf{u} +
\boldsymbol{\nabla}\mathbf{u}^T$}, which is
important for calculation of power dissipation, the energetics of locomotion,
and is also relevant for studies on locomotion in complex, (shear) rate-dependent
fluids. Since the beat-pattern of the nematode is planar, there is no flow in the
$z$-direction in the swimmer mid-plane. However the $z$-derivatives of the
mid-plane velocity will make, in general, non-trivial contributions to the shear
rate. Writing the shear rate in 2D and 3D explicitly, we see
\begin{subequations}	
    \begin{align}
        \dot{\gamma}_{\mbox{\scriptsize{2D}}} &= [2u_x^2 + (u_y + v_x)^2 +
        2v_y^2]^{1/2}, \label{eq:shear_rate2D} \\
        \dot{\gamma}_{\mbox{\scriptsize{3D}}} &= [2u_x^2 + (u_y + v_x)^2 +
2v_y^2 \nonumber \\
&+ 2w_z^2 + (u_z + w_x)^2 + (v_z + w_y)^2]^{1/2}, \label{eq:shear_rate3D}
    \end{align}
\end{subequations}
where $u,v,w$ are the $x,y,z$ components of the velocity field, and subscripts
denote derivative components. 

{Because of the additional terms in the 3D formula, we expect
    calculation of the shear rate from 2D particle tracking flow-fields will
    likely result in a systematic underestimate of the true 3D value.
    Fig.~\ref{fig:shearrate_rel_err}{\it a} shows that this is indeed the case; the
    spatial distribution of the relative percentage error in the simulated shear
    rate calculated with the 2D formula~\eqref{eq:shear_rate2D} compared with
    the true 3D value spatial distribution of the shear rate as calculated by
    the 3D~\eqref{eq:shear_rate3D} is significant, even reasonably far from the
    worm, and is around 30\% close to the worm. Indeed the integrated root mean
    square error in this field is 14\%. Thus, we see that we have significant
errors throughout the field, but particularly close to the worm where
calculations of shear rate are of particular interest.}

{We wish to correct for this error without resorting to Boundary
    Element calculations, and so require estimates of the unknown quantities
    in the 3D formula~\eqref{eq:shear_rate3D}. Since the worm kinematics is
    planar, we have the symmetry $z\rightarrow -z$, and so there
    is no $z$-flow in the mid-plane and the quantities $w_x$ and $w_y$ are zero.
    Furthermore, the quantities $u_z$ and $v_z$ are also zero by this symmetry. These
    observations are confirmed by our numerical simulation, which calculates the
    above quantities to be zero within numerical error. However, the
    $z$-component of the velocity $w$ changes sign through the mid-plane, and
    thus its $z$-derivative $w_z$ makes a significant contribution to the shear
    rate. The 3D formula~\eqref{eq:shear_rate3D} can thus be simplified, 
\begin{equation}
\dot{\gamma}_{\mbox{\scriptsize{pl}}} = [2u_x^2 + (u_y + v_x)^2 +
2v_y^2 + 2w_z^2]^{1/2}. 
\end{equation}
Since the flow is incompressible, we have $\boldsymbol{\nabla}\cdot\mathbf{u} =
0$, so that $w_z = -(u_x + v_y) $, giving the final formula
\begin{equation}
\dot{\gamma}_{\mbox{\scriptsize{pl}}} = [2u_x^2 + 2v_y^2 + (u_y + v_x)^2 + 2(u_x + v_y)^2]^{1/2}, 
\label{eq:shear_rate_planar}
\end{equation}
purely in terms of planar components obtainable via planar particle tracking. Applying the
adjusted formula~\eqref{eq:shear_rate_planar} to our numerical data, the error is
eliminated to within 0.01\% which is attributable to the accuracy of our
numerical scheme.

The principal advantage of this approach is that it can be applied in just as
simple a manner to incompressible non-Newtonian flows, such as shear-thinning and viscoelastic fluids. Such fluids have complex
non-Newtonian constitutive laws, and do not readily admit simple
three-dimensional simulation.

\begin{figure} 
    \begin{center}
        \includegraphics[width=\textwidth]{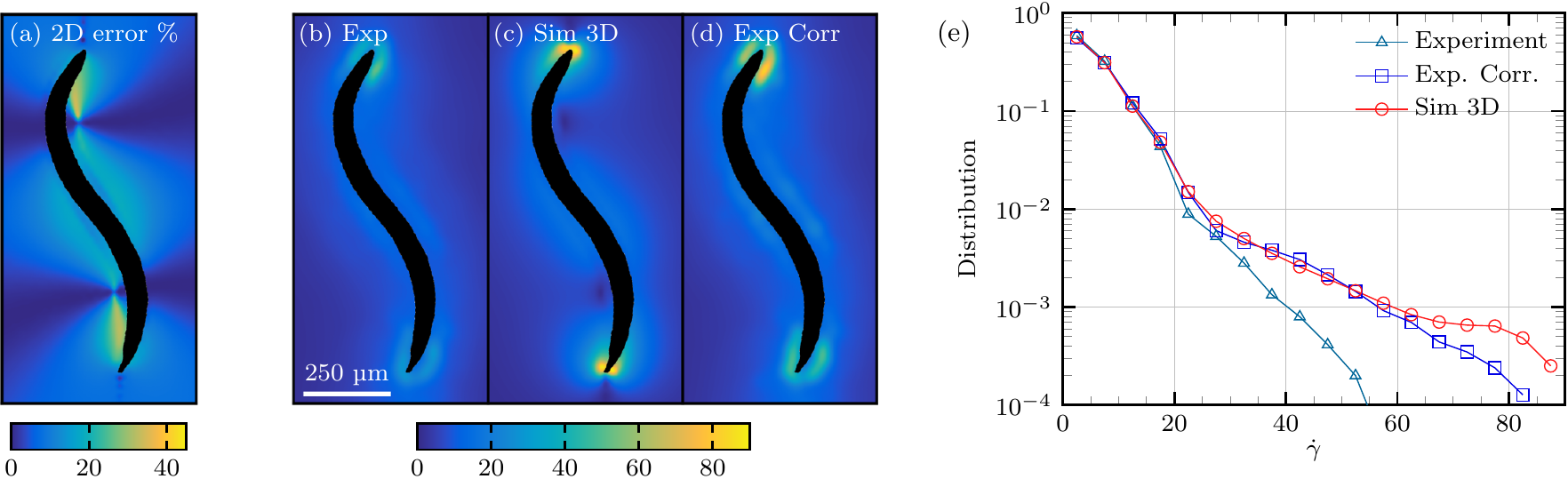}
    \end{center} 
    \caption{ Shear rate error, calculation, and correction: (a) The
        percentage relative error in the shear rate as calculated using the 2D
        formula~\eqref{eq:shear_rate2D} for simulated data, showing large errors
        throughout. (b)-(d) Shear rate field for (b) planar 2D experiments
        calculated with the 2D formula~\eqref{eq:shear_rate2D}, (c) 3D
        simulation calculated with the full formula~\eqref{eq:shear_rate3D}, and
        (d) corrected ``3D'' experiments calculated with the corrected
    formula~\eqref{eq:shear_rate_planar}. (e) Distribution of shear rates for
planar experiments, 3D simulation, and corrected experiments. Note the marked
improvement in similarity between shear rate distribution for the corrected
experiments and that of the full 3D simulation.} 
        \label{fig:shearrate_rel_err} 
\end{figure}

We now directly compute the 2D~\eqref{eq:shear_rate2D} and the estimated
3D~\eqref{eq:shear_rate_planar} shear rate field for our experimental data
(Fig.~\ref{fig:shearrate_rel_err}{\it b}). Figure~\ref{fig:shearrate_rel_err}
shows a comparison between the ``corrected'' 3D shear rate field from
experiments and the 3D shear rate field from simulations \textit{(c,d)}. We find the corrected field from experiments and the simulated 3D field have a strikingly similar structure and an RMS error of only 14\%, compared to an RMS error of roughly 20\% before the correction was applied. We note, however, that there are discrepancies near the head and tail of the worm, where the simulations suggest a slightly higher shear rate ($\approx 10\%$); these regions of high velocity near the swimmer-fluid interface are in locations where we expect particle tracking techniques to pose the greatest challenge.

To quantify the differences between the shear rate fields of the simulations and
experiments, as well as demonstrate the effectiveness of our correction factor
for 2D data, we show the distribution of shear rate for the raw experimental
data, corrected experimental data, and the full 3D simulations
(Fig.~\ref{fig:shearrate_rel_err}{\it e}). We find that a 2D calculation of shear rate using our experimental data underestimates shear rate for values of $\dot{\gamma}>25$~s$^{-1}$ compared to the 3D simulated shear rate field. When the Eq.~\ref{eq:shear_rate_planar} is applied to the experimental data, we capture the same shear rate distribution as the 3D simulations up to $\dot{\gamma}=60$~s$^{-1}$. For larger shear rates ($\dot{\gamma}>60$~s$^{-1}$), we see only minor deviations from the 3D simulations, suggesting that the corrected shear rate field significantly improves our ability to estimate the shear rate magnitude from planar data.

Since the shear rate is calculated using derivatives of the velocity field,
small fluctuations or errors in the velocity field are magnified substantially.
Nonetheless, we obtain good agreement between experiment and simulation, with
errors being confined to regions of high shear rate; importantly, we find that
the application of the analytical correction factor greatly improves our
estimate of the maximum shear rate around the swimmer, which is particularly
useful for swimming applications with non-Newtonian fluids. These results indicate that the local shear rate near a low Reynolds number
swimmer is likely much higher than previously thought, due to contributions in
the third dimension that planar particle tracking velocimetry does not directly measure. As
a result, non-Newtonian effects as a result of locomotion in complex fluids may be much larger than
anticipated. Examples include the role of elastic stretching, measured by the
Weissenberg number ${\mbox{\textit{Wi}}}=\lambda_{E} \dot{\gamma} $ where
$\lambda_{E}$ is the longest relaxation time of the fluid, and shear-thinning
viscosity behavior, measured by the Carreau number ${\mbox{\textit{Cr}}} =
\lambda_{{\mbox{\textit{Cr}}}} \dot{\gamma} $ where
$\lambda_{{\mbox{\textit{Cr}}}}$ is a timescale that represents the onset of
shear-thinning effects; our accounting for the shear rate in the third dimension
therefore suggests that a planar experimental measurement may under-represent
non-Newtonian effects near the body of a the swimmer.}

\section{Conclusion}

In this work, we used theoretical and numerical techniques to improve the
processing of experimentally-obtained particle tracking data, producing smooth
velocity fields quantifying the flow around the swimming nematode
\textit{C.~elegans} in a Newtonian solution. We compared our results with a 3D
boundary element model of the nematode, generated directly from experimentally
obtained nematode kinematics, finding good agreement between numerics and
experiment. 

{ We argued that when calculating derivative flow quantities, the
    only non-trivial out-of-plane component is $w_z$: the $z$-derivative of the
    $z$-flow. This observation was validated by simulation, showing that excluding
$w_z$ when calculating the flow shear rate results in a significant
underestimate. Accounting for the $w_z$ component via incompressibility from 2D
data eliminated this error.}

Our work illustrates how theory may be used to improve experimental measurements
in biological fluid mechanics, and will be directly applicable to investigations
of bio-locomotion in complex fluids. Furthermore, we anticipate that as
technology for the acquisition and processing of experimental flow-fields
continues to improve, these out-of-plane effects will represent a hard barrier
to increasing the accuracy of results, making such techniques increasingly important.

\begin{acknowledgements} T.D.M-J. is supported by a Royal Commission for the
Exhibition of 1851 Research Fellowship and D.A.G. is supported by an NSF
Graduate Fellowship. Funding from the European Union (CIG Grant to E.L.) and
NSF-CBET-1437482 (to P.E.A.) is gratefully acknowledged. {The
    authors would like to thank the anonymous referees for their helpful
comments on the manuscript.}
\end{acknowledgements}


\end{document}